\documentclass[letterpaper]{article}
\begin{document}

\title{ Elastic Scattering and Total Reaction Cross Section \\
 for the $^{6}$He + $^{27}$Al System}
\author{
E. A. Benjamim$^1$, A. L\'epine-Szily$^1$, D. R. Mendes Junior$^1$, 
R. Lichtenth\"aler$^1$,\\
 V. Guimar\~aes$^1$, P. R. S. Gomes$^2$, L. C. Chamon$^1$, 
M. S. Hussein$^1$, A. M. Moro$^3$,\\  A. Arazi$^4$, I. Padron$^{2,5}$,  
J. Alcantara Nu\~nez $^1$, M. Assun\c{c}\~ao$^1$, A. Barioni$^1$,\\ 
O. Camargo Jr.$^1$, R. Z. Denke$^1$,  
P. N. de Faria$^1$, K. C. C. Pires$^1$\\ 
\normalsize $^1$ {\it Instituto de F\'{\i}sica - Universidade de S\~ao Paulo, C.P.66318,05389-970 S\~ao Paulo, Brazil}\\
\normalsize $^2$  {\it Instituto de F\'{\i}sica - Universidade Federal Fluminense, Niter\'oi,R.J.,24210-340,  Brazil}\\
\normalsize $^3$ {\it Departamento de FAMN, Universidad de Sevilla, Apdo 1065, E-41080, Sevilla, Spain}\\ 
\normalsize $^4$ {\it Laboratorio Tandar, Departamento de F\'{\i}sica, Comis\'{\i}on Nacional de Energ\'{\i}a}\\
\normalsize {\it At\'omica, Av. del Libertador 8250, (1429), Buenos Aires, Argentina}\\
\normalsize $^5$ {\it CEADEN, P.O. Box 6122, Havana, Cuba}}
\maketitle

\begin{abstract}
The elastic scattering of the radioactive halo nucleus $^{6}$He on $^{27}$Al 
target was measured at four energies close to the Coulomb barrier using the
 RIBRAS (Radioactive Ion Beams in Brazil) facility. The S\~ao Paulo Potential 
(SPP) 
was used and its diffuseness and imaginary strength were adjusted to fit 
the elastic scattering angular distributions.
Reaction cross-sections were extracted from the optical model fits. The 
reduced reaction cross-sections of $^{6}$He on $^{27}$Al are similar to 
those for stable, weakly bound projectiles as $^{6,7}$Li, $^9$Be 
and larger than stable, tightly bound projectile as $^{16}$O on $^{27}$Al.
 \end{abstract}

\vspace{0.2cm}

Reactions induced by halo nuclei are currently one of the main subjects 
in Nuclear Physics, with interest in nuclear structure, reactions, 
astrophysics  and production of superheavy elements. 
 Nowadays reliable measurements of elastic, inelastic and transfer 
cross sections of unstable projectiles are possible due to new 
Radioactive Nuclear Beam (RNB) facilities. Nuclei that have small separation 
energies have large probability of 
breaking-up when the colliding nuclei approach each other. In this case
their interactions convert potential and kinetic energy into relative kinetic
 energy between their fragments.  Many efforts have been made, both 
theoretical and experimental, to investigate the effect of the 
break-up on the fusion cross section \cite{can05}. 
 However most of the experiments in the 
field of fusion have been carried out with intense beams of stable 
weakly bound nuclei ($^6$Li, $^7$Li, $^9$Be), which have respectively 
separation energies of
$S_{\alpha}$=1.48 MeV,  $S_{\alpha}$=2.45 MeV and $S_{n}$=1.67 MeV, 
 larger than those for halo radioactive nuclei.
 Nevertheless important features of the reactions induced
 by halo nuclei are not present in those reactions. As an example, 
we mention that halo nuclei have larger nuclear radii, which corresponds to a
lowering of the Coulomb barrier. Also, important couplings to the soft dipole 
resonance that may be present in the case of halo nuclei with quite different 
proton and neutron distributions, do not occur for normal nuclei. 
For the 2 neutron halo nuclei with Borromean nature  
(while the 3 body (core + 2n) system is bound, any of the 2 body 
subsystems (core + 1n) or (2n) is unbound) the $^{11}$Li and $^{6}$He 
are good examples. The correlation between the 2 neutrons of the halo
is also important feature
as recent measurements on the Coulomb dissociation of $^{11}$Li have 
demonstrated \cite{nak06}. Strong low-energy (soft) E1 excitation was observed,
 peaked at about Ex=0.6 MeV with B(E1)=1.42(18) e$^2$ fm$^2$. The spectrum 
is reproduced well by a three-body model with a strong two-neutron correlation.
Similar features have been observed for other two-neutron 
Borromean nuclei $^{6}$He \cite{aum99} and $^{14}$Be \cite{lab01} and 
possibly for the two-proton Borromean nucleus $^{17}$Ne \cite{kan03}.
  Break-up effects are also expected to play an important role in the 
scattering mechanism, affecting the imaginary part of the interaction 
potential. So far, conflicting predictions have been made about 
whether the fusion
 of weakly bound nuclei is enhanced or hindered owing to the strong coupling 
to the break-up channel \cite{hus92,tak93,das94,can95,hag00,dia02}. 
The role of the Coulomb and nuclear break-up and their interferences 
on the fusion process
is a subject of great interest \cite{can95,dia02,gui00}.
 One of the important questions is whether the 
effect of the break-up is essentially to increase the total reaction 
cross section, instead of affecting the fusion cross section. 
 Depending on whether the break-up 
is dominated by the Coulomb or nuclear interactions, different effects on 
the fusion cross section could result. 
Therefore, 
it is essential to investigate the dependence of the break-up and total 
reaction cross sections on the break-up threshold energy of different 
projectiles and on the target mass.

So far, reactions induced by halo nuclei have been studied mostly on heavy 
targets, such as $^6$He + $^{209}$Bi \cite{agu00,kol98,you05}, 
$^6$He + $^{238}$U \cite{tro00,raa04}, and $^{11}$Be + $^{209}$Bi 
\cite{sig99}, although recently experiments have been performed with 
$^6$He projectile on 
medium mass targets, namely $^{64}$Zn \cite{pie03} and $^{65}$Cu 
\cite{nav04}. For heavy systems, where the Coulomb break-up predominates,
 very large total reaction 
cross sections have been obtained at energies 
 around and below the Coulomb barrier
due to the intense long range Coulomb field. 
In this case the transfer + break-up cross sections were found to account for 
most of the total reaction cross section.
 For the medium mass targets, where the 
Coulomb field is not so intense, transfer and break-up cross sections were 
found to be much more important than the fusion cross section 
at energies above 
the Coulomb barrier and the total reaction cross sections were much 
larger than for 
$^{4}$He or $^{6}$Li projectiles \cite{pie03,nav04}. 
In a comparison 
between total reaction cross sections for $^6$He, $^6$Li, $^7$Li, $^9$Be, 
$^{16}$O projectiles on $^{64}$Zn target, Gomes et al \cite{gom04} have 
shown that the reaction cross section is largest for the $^{6}$He 
(halo nucleus with threshold energy of 0.975 MeV), followed by the 
group of the three stable weakly bound projectiles, and finaly the tightly 
bound $^{16}$O projectile produces the smallest total reaction cross section. 

In this work we have investigated the behavior of the total reaction 
cross section 
of the $^{6}$He + $^{27}$Al system, 
and compared it with the available data for the 
$^9$Be + $^{27}$Al \cite{mar05,go04}, $^6$Li + $^{27}$Al, $^7$Li + $^{27}$Al
\cite{fig05} and $^{16}$O + $^{27}$Al systems \cite{cre79}. 
For the $^6$He + $^{27}$Al system the  contribution from 
the nuclear break-up should be predominant not just 
due to the weaker Coulomb field, but also due 
to the long range nuclear interaction of the two neutron halo 
projectile $^6$He with the target.
         For $^{9}$Be + $^{27}$Al, despite the relatively weak  Coulomb
 field, it was observed \cite{mar05}  that the transfer plus break-up 
processes still have a significant cross section. The study of the elastic 
scattering and transfer + break-up processes for the $^{6,7}$Li + $^{28}$Si
 systems \cite{pak03} also shows the importance of the nuclear break-up 
process for  stable weakly bound nuclei on a light mass target.

The study of light systems with very weakly bound and neutron 
rich exotic nuclei is particularly interesting, since there are reactions of
great astrophysical interest involving these nuclei. As an example, in the
 case of 
inhomogeneous distribution of protons and neutrons following the Big Bang
not only stable light elements but also proton and neutron
rich short-lived  elements such as $^6$He, $^7$Be, $^8$B and $^8$Li would be
 present
in the early universe.  These, short lived, radioactive nuclei could thus 
bridge  the
A=8 mass gap and heavier  elements would then be synthetized. Reactions 
involving
light unstable nuclei would be present also in the Type-II supernovae, 
neutron stars
and in massive stars.  Besides the triple alpha capture, 
the $\alpha$-recombination and
the bridge of mass 5 and 8 in the  beginning of r-process in a Type-II  
supernova could
be given via alternate three-body reactions or sequential  capture reactions 
such 
as $^4$He(2n,$\gamma$)$^6$He(2n,$\gamma$)$^8$He \cite{gor95}.  In this case, 
the two neutron
capture  reaction cross sections on $^4$He and $^6$He depend strongly  on the
pronounced halo structure of the $^6$He and $^8$He compound nuclei.

 In a recent paper Milin et al. \cite{mil04} have presented elastic and 
 inelastic scattering as well as 2n transfer reaction angular distributions 
for the  $^6$He + $^{12}$C 
system, measured at E$_{lab}$ = 18 MeV, which is well above the 
Coulomb barrier 
(3 MeV for the $^6$He + $^{12}$C system). The main goal of their work 
was the spectroscopic investigation of stretched neutron configurations
in the $^{14}$C final nucleus.

This is the first paper reporting on experiments with radioactive beams 
using the recently installed RIBRAS facility, at S\~ao Paulo \cite{lic03}. 
This facility 
is installed at the Pelletron Laboratory of the University of S\~ao Paulo 
and extends the capabilities of the original Pelletron Tandem Accelerator 
of 8MV terminal voltage (8UD) by 
producing secondary beams of unstable nuclei. 
The most important components of this facility are the two new
 superconducting solenoids with 6.5 T maximum central field and a 30 cm
 clear warm bore, which corresponds to a maximum angular acceptance 
in the range of 
2 deg$\leq\theta\leq$15 deg. The solenoids 
are installed on the 45B beam line of the Pelletron Tandem. Thus the 
actual system is similar to the TWINSOL facility at Notre Dame University
 \cite{kol89,bec03}.
      The presence of the two magnets is very important to produce pure 
secondary beams. However in the present work only the first solenoid was 
used. When using only one solenoid the secondary beam still has some 
contaminants easily identified in elastic scattering experiments. 
    
The production system consists of a gas cell, 
mounted in a ISO chamber with a $2.2 \mu m$ Havar
 entrance window and a  $^9Be$ vacuum tight exit window $16 \mu m$ thick, 
which plays the role of the  
primary target and the window of the gas cell at the same time.
The gas inside the cell has the purpose of cooling the Berilium foil
heated by the primary beam and can also be used as production target. 
A few centimeters behind the gas cell there is a tungsten rod
 of $2.4$cm diameter with a cilindrical hole of $1$cm diameter
per $2$cm length, to stop and collect the primary beam particles. 
In front of the 
W beam stopper there is a circular colimator which is polarized to
 a negative voltage of about $-250$ Volts
 in order to supress the secondary electrons emitted 
from the cup.
 The experimental set-up, with the production target, the W beam stopper,
 the magnets and the scattering chamber, with secondary target and 
detectors is presented in Fig. 1. 

The $^6$He secondary beam was produced by the $^9$Be($^7$Li,$^6$He)$^{10}$B 
reaction with Q = - 3.38 MeV, by impinging about 300 nAe
primary beam of $^7$Li on the 16 $\mu$m thick $^9$Be production target.
The primary beam is stopped in a Faraday cup, constituted by an isolated 
Tungsten rod which stops all particles in the angular region from 0 to 2 
degrees and where the primary beam intensities were integrated. 
    The reaction products of the secondary beam were detected using 
two $\Delta$E-E Si 
telescopes, with detector thicknesses of 22-150 $\mu$m and 50-150 $\mu$m 
respectively, with antiscattering collimators separated by 4 cm in front of 
the $\Delta$E detectors. The distance from the secondary target to the 
collimator, which limited the solid angle to 15 msr,
 was 8 cm and the angular opening 
of the collimators was about $\pm 2.6$ degrees.
The secondary beam was not pure, reminiscence of the primary beam was 
detected at zero degrees in the 2+ charge state, as well as $^4$He, $^3$H 
and protons were transmitted with the appropriate energy through the first
 solenoid. 
The secondary targets used in this experiment were a $^{27}$Al target of 
7.2 mg/cm$^2$ and a $^{197}$Au target of 5 mg/cm$^2$.
   
The secondary beam intensities were calculated by assuming pure Rutherford 
scattering of the $^6$He on the gold target. 
From this intensity and from the integrated charge of the primary beam 
intensity for every run one can derive the production efficiency which is 
the ratio of secondary to 
primary beam intensities. As this efficiency does not depend on the target 
but only on the energy, it was used for the determination of the 
secondary beam intensities on Aluminum targets allowing the calculation of the 
absolute cross sections. 
   
The effective solid angles and the average detection angles were 
determined by using a Monte Carlo simulation, which took into 
account the collimator size, 
the secondary beam spot size on the secondary target ($\phi$=4 mm), 
the secondary 
beam divergence (1.5-3.5 degrees, was limited by collimators following the 
primary target) and the angular distribution at forward angles in the 
detector, which modifies the average detection angles. The total angular
 uncertainty of $\pm$ 3.2 degrees was 
calculated with the Monte Carlo simulation and includes the beam spot size, 
the secondary beam divergence and angular straggling in the target.
      
We have used the primary $^7$Li beam at the four different incident energies 
19.0, 20.5, 21.0 and 22.0 MeV. The secondary beam energies were calculated 
by energy losses and confirmed by the energy measurement in the Si 
telescope, calibrated with alpha particles of a radioactive $^{241}$Am source.
The $^6$He secondary beam energies in the middle of the $^{27}$Al secondary 
target were respectively 9.5, 11.0, 12.0  and 13.4 MeV, while the value of the 
Coulomb barrier in the laboratory system for the $^6$He +$^{27}$Al system 
is about 8.0 MeV. The total energy loss of the $^6$He beam in the thick Al 
target is about 3.2 MeV. After passing through the
thick 27Al target, the 6He beam has lost 1.6MeV on its way 
to the middle of the target and 1.6MeV on its way out. 
These energy losses also apply roughly for the scattered particles. The 
energy resolution is determined by three components: the energy 
resolution of the $^6$He beam,
the energy straggling in the target and the kinematical broadening of 
the scattered particles. The energy resolution (FWHM) with the thick Al target
 is the the full width  of the elastic peak in the 
energy spectrum and its value is about 400 keV, thus sufficient to 
separate the inelastic excitation of the first excited state of 
$^{27}$Al at 843 keV.
  The $^6$He secondary beam intensity at these energies was about  
0.7$\times 10^5$ pps/($\mu Ae$ of $^7$Li).
  

  The results for the angular distributions are shown in Fig. 2.
 The error bars in the cross 
sections of the angular distributions are due to the 
statistical errors. The error bars in the scattering angles 
are of $\pm$ 3.2 degrees and were 
calculated with the Monte Carlo simulation.

The elastic scattering cross sections of the angular distributions were 
reproduced by  optical model calculations obtained with 
the S\~ao Paulo Potential (SPP)
 \cite{cha02,alv03}, which is 
a folding optical potential which takes nonlocal effects into account. 
The imaginary 
part of the potential has the same form factor as the real part and 
the only free 
parameters were the normalization of the imaginary potential, $N_I$ 
and $a$, the 
diffuseness of the nuclear density of the projectile.

 We have 
fitted the angular 
distributions by minimizing the $\chi^2$. In this procedure, 
$N_I$ and $a$ were varied, but we have assumed that a common 
diffuseness value should be assumed for all energies.
The best value of the diffuseness 
was $a$=0.56(2) fm. 
 The best values $N_I$ were 0.8(5), 0.8(4), 0.6(4) and 0.7(5) 
for the energies 9.5, 11.0, 12.0  and 13.4 MeV, respectively.  The errors 
in the parameters are determined from  the $\chi^2$ parabola, 
at the value of $\chi^2 = \chi^2_{min}+1$. 
The reaction cross sections 
were calculated using the best fit optical potentials, with the best common
diffuseness $a$ = 0.56 fm and the best value of $N_I$ for each energy.
 Their uncertainty comes from the quoted uncertainties in the 
optical potential, namely in the diffuseness $a$
and in the imaginary normalization factor $N_I$. The large error in the  
reaction cross sections  ($\pm $ 100 mb), comes mainly from the large
 error in the imaginary normalization factor $N_I$. 
If we use the best common $N_I$ for all angular distributions, 
$N_I$=0.65(20), the error in $N_I$ is reduced. 
  The best diffuseness, $a$=0.56(2)fm  corresponds to the diffuseness 
of the matter distribution of the $^6$He nucleus and is the convolution 
of the point nucleon distribution of $^6$He with 
the intrinsic matter distribution of the nucleon. From the 
deconvolution we obtained the diffuseness of the point nucleon distribution 
of $^6$He as being 0.52(2) fm.
This diffuseness is much larger than 
the diffuseness of $^4$He, which is about 0.3 fm, deduced in similar 
manner \cite{gas03}. The  $^6$He density obtained in the present 
work is compatible with other experimental evidences \cite{alk97,gas03}
and with theoretical calulations \cite{zhu93,alk96}.

For the other systems, as 
$^9$Be + $^{27}$Al \cite{mar05,go04}, $^6$Li + $^{27}$Al, $^7$Li + $^{27}$Al
\cite{fig05} and $^{16}$O + $^{27}$Al \cite{cre79}, the elastic 
scattering angular distributions available were analysed using the 
same procedure employed for the $^6$He + $^{27}$Al system. 
All elastic scattering angular
 distributions were fitted using the SPP,  
allowing for the variation of $N_I$ and density diffuseness $a$.
 We have 
obtained different values of $a$ for the different 
systems, as $a$=0.53 fm for the $^{16}$O, 0.56 fm for the $^7$Li, and 
0.58 fm for the $^6$Li and $^9$Be. All systems were analysed using 
the same procedure for consistency. All reaction cross 
sections were obtained from the fit of the elastic scattering 
angular distributions 
with SPP.
The reaction cross sections determined in this way are presented 
in the Table I. 

In order to compare total reaction cross sections for different
 systems we used 
the procedure suggested in ref \cite{gom05}, where the cross sections 
are divided by $({A_p}^{1/3} + {A_T}^{1/3})^2$ and the center of mass energy 
by $Z_p Z_T$ / $({A_p}^{1/3} + {A_T}^{1/3})$, where $Z_p$ ($Z_T$) 
and $A_p$ ($A_T$) 
are the charge and mass of the projectile (target), respectively. In this way, 
the geometrical effects are removed and the eventual anomalous 
values of the reduced 
radii $r_0$, which should be related to the physical processes to be 
investigated, are not washed out.

Figure 3 shows the results of the reduced total reaction cross sections,
 $\sigma_{R}^{red}$,
for the halo 
$^6$He, the stable weakly bound $^9$Be, $^{6,7}$Li and the 
tightly bound $^{16}$O projectiles on the same $^{27}$Al target. 
One can observe that the smallest cross section is for the 
tightly bound projectile $^{16}$O + $^{27}$Al, for which the break-up 
process is not expected 
to occur. The values of $\sigma_{R}^{red}$ for the weakly bound 
stable nuclei on $^{27}$Al are similar 
to each other and are larger than that for the $^{16}$O. Their 
uncertainties are of the 
order of 2-4\%, due to the difference in the reaction 
cross section when different potentials are used, which 
give very similar quality in the fits. 
The values of $\sigma_{R}^{red}$ for the halo nucleus $^6$He, which  
has the smallest break-up 
threshold energy and has the largest uncertainties, is seemingly similar 
to $\sigma_{R}^{red}$ of the weakly bound stable nuclei. 
However the relation employed to obtain $\sigma_{R}^{red}$ is entirely 
based on the use of the radius to mass number relation, $R$ = $r_0A^{1/3}$, 
which is not quite appropriate for halo nuclei such as $^6$He.


We have estimated the nuclear break-up cross-section by extending 
the closed 
formalism of Frahn for heavy ion inelastic scattering \cite{fra76}, 
based on the
adiabatic Austern-Blair theory \cite{aus65}, to the break-up case by 
considering the
latter as directly related to the multipole polarizability of the weakly bound
nucleus \cite{hus06}. If only dipole and quadrupole terms are considered we 
find, 

\begin{equation}
\sigma_{break-up} \cong \Big[ (\delta_{1})^2  \Big( \frac{3}{2} \frac{\Delta R}{R} \Big)^2 + (\delta_{2})^2  \Big] \sum_l (2l+1) \mid \frac{d S_{N}(l)}{dl}\mid ^2
\end{equation}

\noindent
where, $S_{N}(l)$ are the scattering matrix elements of the $^6$He+$^{27}$Al 
collision calculated with the best fit optical potential, 
$\delta_i $ is the $i^{th}$ multipole deformation length and 
$\Delta R \equiv R_n - R_p $ is the difference in the neutron and proton 
rms radii, while $R$ is the rms matter radius. For $^6He$, 
$\Delta R$ = 0.61 $\pm$ 0.21 fm and $R$ = 2.3 $\pm$ 0.07 fm \cite{alk97}. 
In obtaining Eq.(1) we have ignored in the dipole contribution  a term 
proportional to $d^2S_N(l)/dl^2$. The dipole, $\delta_1$, and quadrupole, 
$\delta_2$, deformation lengths were deduced in \cite{aum99} to be roughly 
equal and given by $\delta_1 \sim \delta_2$ = 1.8 fm. Our estimate, based on 
Eq.(1), of the nuclear break-up cross section for $^6$He+ $^{27}$Al is about 
150 mb at E$_{lab}$ = 10 MeV, of which the dipole contributions is about 
30 mb while the contribution of the quadrupole is 120 mb.  The estimate of 
$\sigma_{break-up}$ above does not include the Coulomb break-up which, 
if added,  would give rise to a larger total break-up cross-section. 
Preliminary CDCC calculations \cite{moro06} where the two  halo neutrons 
in $^6$He are treated as one dineutron entity, give very similar numbers as
 Eq.(1). The quadrupole deformation lengths of $^6$Li, $^7$Li and  $^9$Be were
 calculated from their quadrupole moments \cite{rag89} and nuclear rms matter
 radii \cite{tan95} and their values are respectively 0.014, 0.73 and 0.70 fm.
The dipole deformation of the $^{6}$Li, in the $\alpha$-d cluster 
model is zero. Thus the nuclear break-up cross sections, calculated by Eq. 1, 
 of the weakly bound $^6$Li, $^7$Li and  $^9$Be nuclei on $^{27}$Al 
are not more than 20\% of the corresponding  break-up cross section 
 for the $^6$He + $^{27}$Al system.
 
Equation (1) was derived assuming the validity of the adiabatic approximation, 
where the Q-value is ignored. This is the basis of the Austern-Blair formula.
It should give an upper limit to the cross section. It is true that the 
Austern-Blair theory was originally employed for inelastic excitations 
of collective states. We do not see any reason why not using it for 
break-up considered as inelastic excitation into the continuum through 
dipole, quadrupole, etc, transitions. The nuclear break-up cross sections of
 $^6$Li, $^7$Li and  $^9$Be, obtained with Eq. (1) are less than 20\% of that of $^6$He.
This is a correct statement since the deformation lengths are much smaller and, 
further, the Q-values, not included in Eq. (1) are larger.

In summary, we report in this Letter the first results obtained with the new 
radioactive beam facility RIBRAS at S\~ao Paulo, with the measurement 
of elastic scattering between 
$^6$He and the light $^{27}$Al target, at four energies 
slightly above the Coulomb 
barrier. The derived total reaction cross sections were 
compared with other systems, 
the stable weakly bound $^9$Be, $^{6,7}$Li and the tightly bound 
$^{16}$O projectiles on the same target. The reduced reaction cross sections 
for the 2 neutron Borromean halo nucleus $^6$He on $^{27}$Al are similar 
to other weakly bound 
stable systems within the error bars. This indicates  
that for light systems the effects of the halo on the 
reaction cross sections could be much smaller than 
the effect observed in heavier sytems. Certainly higher 
quality data are needed to observe the effect of the Borromean 
nature of $^6$He on the reaction cross section at low energies.

\vspace{0.5cm}
Acknowledgements\\ 
The authors wish to thank the Funda\c{c}\~ao de Amparo \`a Pesquisa do Estado 
de S\~ao Paulo (FAPESP) and the Conselho Nacional de Desenvolvimento 
Cient\'{\i}fico e Tecnol\'ogico (CNPq) for financial support (FAPESP 2001/06676-9 
and 2003/10099-2).\\

\newpage
{\tiny{\bf
\begin{center}
\begin{tabular}{r c c c c c c c c }\hline\hline
 {System} & {E$_{lab}$} & $\sigma_R${(mb)} & {E$_{cm}^{red}$} & $\sigma_R^{red}$\\
\hline
& \\
$^{6}$He + $^{27}$Al& 9.5  & 1110(90) & 1.44  & 48(4) \\
& 11.0 & 1257(100) & 1.67 & 54(4)  \\                                                                                                      
& 12.0 & 1300(100) & 1.82 & 56(4) \\
& 13.4 & 1390(100) & 2.03 & 60(4) \\
& \\
\hline
& \\
$^{6}$Li + $^{27}$Al & 7.0  & 113(3) & 0.71  & 5(1) \\
& 8.0 & 320(19) & 0.81 & 14(1)  \\                 
& 10.0 & 625(4) & 1.01 & 27(2) \\
& 12.0 & 913(55) & 1.21 & 39(2) \\
& \\
\hline
& \\
$^{7}$Li + $^{27}$Al & 6.0  & 51(2) & 0.60  & 0.52(6) \\
 & 7.0  & 153(4) & 0.70  & 2.10(6) \\
& 8.0 & 226(7) & 0.80 & 9.4(3)  \\                 
& 9.0 & 364(11)  & 0.90 & 15.1(4) \\
& 10.0 & 536(16) & 1.00 & 22.0(7) \\
& 11.0 & 728(22) & 1.10 & 30.2(9) \\
& 12.0 & 840(25) & 1.20 & 35(1) \\
& 14.0 & 1055(32) & 1.40 & 44(1) \\
& 16.0 & 1240(37) & 1.60 & 51(2) \\
& 18.0 & 1294(39) & 1.80 & 54(2) \\
& \\
\hline
& \\
$^{9}$Be + $^{27}$Al & 12.0  & 380(11) & 0.88  & 14.7(4) \\
& 14.0 & 583(35) & 1.03 & 23(1)  \\                 
& 18.0 & 950(57)  & 1.32 & 37(2) \\
& 22.0 & 1250(75) & 1.61 & 48(2) \\
& 25.0 & 1400(84) & 1.83 & 54(2) \\
& 28.0 & 1520(46) & 2.05 & 59(2) \\
& 32.0 & 1670(100) & 2.34 & 65(2) \\
& \\
\hline
& \\
$^{16}$O + $^{27}$Al & 30.0  & 428(8) & 1.00  & 14.0(3) \\
& 35.0 & 678(13) & 1.17 & 22.3(4)  \\                 
& 40.0 & 831(17) & 1.33 & 27.3(5) \\
& 45.0 & 979(20) & 1.50 & 32.1(6) \\
& 45.6 & 994(20) & 1.52 & 32.6(6) \\
& \\
\hline\hline
\end{tabular}
\end{center}
}}
{\vspace{0.5cm}

\noindent
 Table I. The reaction cross 
sections were obtained from the fit of the elastic scattering 
angular distributions of all systems 
with the S\~ao Paulo Potential (SPP)  as explained in the text.
}
{\vspace{2.5cm}



\newpage
\begin{figure}
\vspace*{6.0cm}
\hspace*{-1.0cm}
\includegraphics{Canaliz_solen_novo.eps}
\caption{ The experimental set-up, with the production target, 
the W beam stopper,
 the magnets and the scattering chamber, with secondary target and 
detectors. See text for details.}
\end{figure}

\begin{figure}
\vspace*{16.5cm}
\hspace*{2.5cm}
\includegraphics{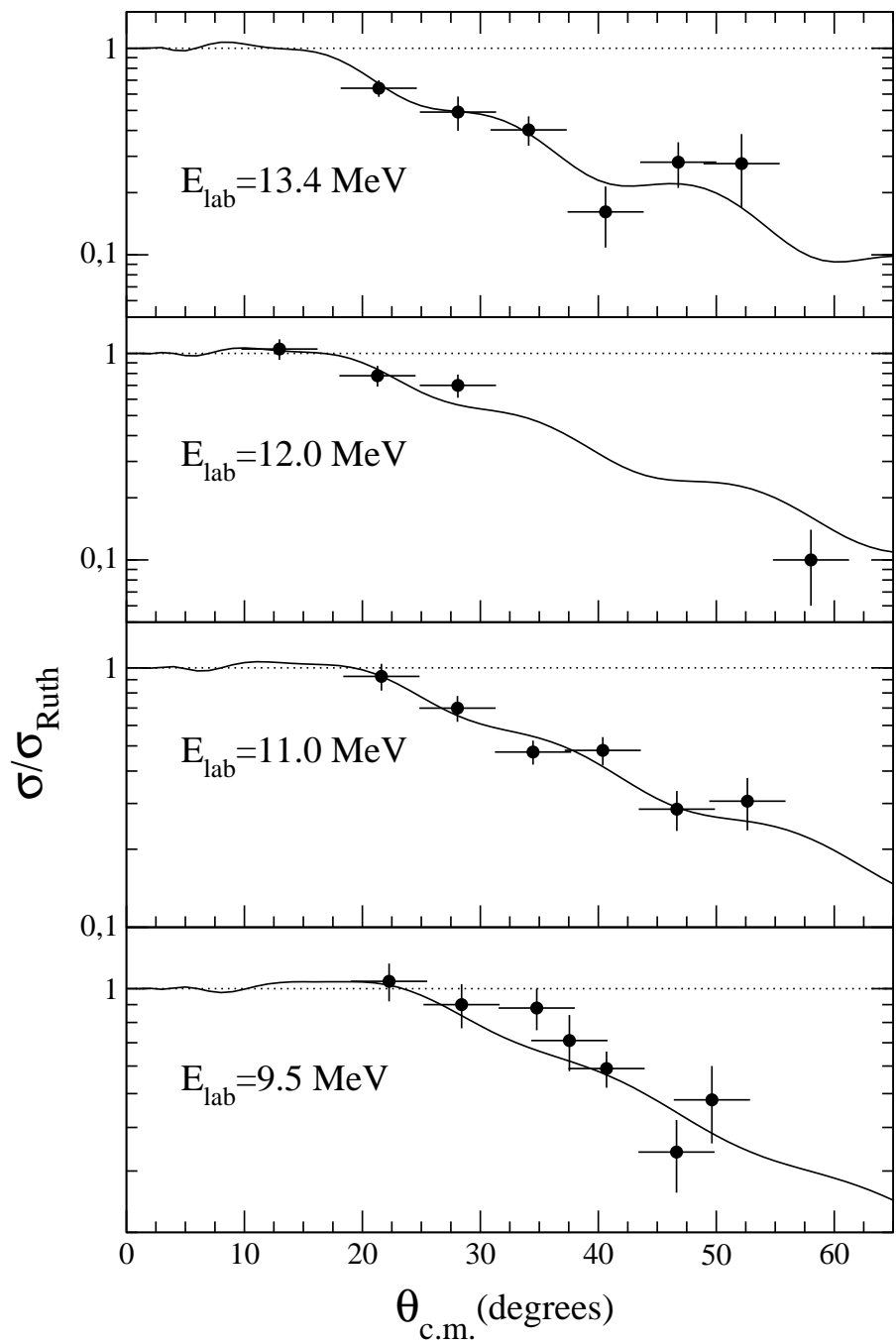}
\noindent
\caption{The  elastic scattering  angular distributions 
measured for the $^6$He+$^{27}$Al system 
together with best fits obtained with the SP potential (SPP). 
The diffuseness of the matter density used in the SPP was $a$=0.56 fm.}
\end{figure}

\newpage
\begin{figure}
\vspace*{-4.0cm}
\hspace*{0.0cm}
\includegraphics{Sig_Red_Artigo_Sem_CDCC.ps}
\vspace*{13.0cm}
\noindent
\caption{The reduced reaction cross sections for the $^6$He+$^{27}$Al system 
obtained in this work together with reduced reaction cross sections of some 
weakly bound stable projectiles and of $^{16}$O on $^{27}$Al.} 

\end{figure}

\begin{thebibliography}{References}
\bibitem{can05}
L.F. Canto {\it et al.,} Phys. Rep. {\bf424}, 1 (2006).
\bibitem{nak06}
T. Nakamura {\it etal.,} Phys. Rev. Lett. {\bf 96}, 252502 (2006).
\bibitem{aum99}
T. Aumann {\it et al.,} Phys. Rev. C {\bf 59}, 1252 (1999). 
\bibitem {lab01}
M. Labiche {\it et al.,}  Phys. Rev. Lett. {\bf 86}, 600 (2001).
\bibitem{kan03}
R. Kanungo {\it et al.,} Phys. Lett. B {\bf 571}, (2003) 21.
\bibitem{hus92}
M.S. Hussein {\it et al.,} Phys. Rev. C {\bf 46}, 377 (1992).
\bibitem{tak93}
N. Takigawa {\it et al.,} Phys. Rev. C {\bf 47}, R2470  (1993).
\bibitem{das94}
C. H. Dasso, A. Vitturi, Phys. Rev. C {\bf 50}, R12 (1994).
\bibitem{can95}
L. F. Canto {\it et al.,} Phys. Rev. C {\bf 52}, R2848 (1995), Nucl. Phys. 
{\bf A589}, 117 (1995). 
\bibitem{hag00}
K. Hagino {\it et al.,} Phys. Rev. C {\bf 61}, 037602 (2000).
 \bibitem{dia02}
A. Diaz-Torres, I. J. Thompson, Phys. Rev. C {\bf 65}, 024606 (2002).
 \bibitem{gui00}
V. Guimar\~aes {\it et al.,} Phys. Rev. Lett. {\bf 84}, 1862 (2000).
\bibitem{agu00}
E.F. Aguilera {\it et al.,} Phys. Rev. Lett. {\bf 84}, 5058 (2000); 
Phys. Rev. C {\bf 63}, 061603(R) (2001).
\bibitem{kol98}
 J.J. Kolata {\it et al.,} Phys. Rev. Lett. {\bf 81}, 4580 (1998).
\bibitem{you05}
P.A. De Young {\it et al.,} Phys. Rev. C {\bf 71}, 051601(R) (2005).
\bibitem{tro00}
 M. Trotta {\it et al.,} Phys. Rev. Lett. {\bf 84}, 2342 (2000).
 \bibitem{raa04}
R. Raabe {\it et al.,} Nature {\bf 431}, 823 (2004).
\bibitem{sig99}
C. Signorini {\it et al.,} Europ. Phys. Journ. A {\bf 5}, 7 (1999).
 \bibitem{pie03}
A. Di Pietro {\it et al.,} Europhysics Letters {\bf 64}, 309 (2003); 
 Phys. Rev. C {\bf 69}, 044613 (2004).
\bibitem{nav04}
A. Navin {\it et al.,} Phys. Rev. C {\bf 70}, 044601 (2004).
\bibitem{gom04}
P.R.S. Gomes {\it et al.,} Phys. Lett. B {\bf 601}, 20 (2004).
 \bibitem{mar05}
G. V. Marti {\it et al.,} Phys. Rev. C {\bf 71}, 027602 (2005).
\bibitem{go04}
 P. R. S. Gomes {\it et al.,} Phys. Rev. C {\bf 70}, 054605 (2004).
\bibitem{fig05}
J. M. Figueira {\it et al.,} Phys. Rev. C {\bf 73}, 054603 (2006). 
\bibitem{cre79}
E. Crema,  Masters degree thesis USP, 1979 , unpublished
\bibitem{pak03} 
 A. Pakou {\it et al.,} Phys. Rev. Lett {\bf 90}, 202701 (2003); 
Phys. Lett B {\bf 556}, 21 (2003);
Phys. Rev. C {\bf 69}, 054602 (2004); Phys. Rev. C {\bf 71}, 064602 (2005).
\bibitem{gor95}
J. G\"orres {\it et al.,} Phys. Rev. C {\bf 52}, 2231 (1995).
\bibitem{mil04}
M. Milin {\it et al.,} Nucl. Phys. {\bf A730}, 285 (2004). 
\bibitem{lic03}
R. Lichtenth\"aler {\it et al.,} Eur. Phys. J. A {\bf 25},s01,733 (2005); 
Nucl. Phys. News {\bf 15} (3), 25 (2005).
\bibitem{kol89}
J.J.Kolata {\it et al.,} Nucl. Instr. Meth. B {\bf 40/41}, 503 (1989).
\bibitem{bec03}
F. D. Becchetti {\it et al.,} Nucl. Instrum. and Meth. in Phys. Res.
 A {\bf 505}, 377 (2003).
\bibitem{cha02}
L. C. Chamon {\it et al.,} Phys. Rev. C {\bf 66}, 014610 (2002).
\bibitem{alv03}
 M.A. G. Alvarez {\it et al.,} Nucl. Phys. {\bf A723}, 93 (2003).
\bibitem{zhu93}
M. V. Zhukov {\it et al.,} Nucl. Phys. {\bf A552}, 353 (1993).
\bibitem{alk96}
J. S. Al-Khalili {\it et al.,} Phys. Rev. C {\bf 54}, 1843 (1996).
\bibitem{gas03}
L. R. Gasques {\it et al.,} Phys. Rev. C {\bf 67}, 024602 (2003); 
Phys. Rev. C {\bf 67}, 067603 (2003).
\bibitem{alk97}
G. D. Alkhazov, {\it et al.}, Phys. Rev. Lett. {\bf 78}, 2313 (1997).
\bibitem{gom05}
P. R. S. Gomes {\it et al.,} Phys. Rev. C {\bf 71}, 017601 (2005).
\bibitem{fra76}
W. E. Frahn, Nucl. Phys. {\bf A272}, 413 (1976).
\bibitem{aus65}
N. Austern and J.S.Blair,  Annals of Physics {\bf 33}, 15 (1965).
\bibitem{hus06}
M. S. Hussein, to be published.

\bibitem{moro06}
A.M. Moro, {\it et al.}, to be published.
\bibitem{rag89}
P. Raghavan, Atomic Data and Nucl. Data Tables {\bf 42}(2), 189 (1989).
\bibitem{tan95}
I. Tanihata, {\it et al.}, Phys. Rev. Lett. {\bf 55}, 2676 (1985).
\end{thebibliography}
\end{document}